\documentclass[a4paper,aps,pra,preprintnumbers]{revtex4}
\usepackage[utf8]{inputenc}
\usepackage[english]{babel}
\usepackage[T1]{fontenc}
\usepackage{amssymb,amsfonts,amsmath,mathtext,enumerate,float,dsfont}
\usepackage{graphics,graphicx,epsfig,epstopdf}
\usepackage{caption}
\usepackage{cmap}
\usepackage{multirow}
\usepackage{indentfirst}
\usepackage[usenames]{color}
\usepackage{amsthm}
\usepackage{xcolor}

\begin{document}

\title{Bosonic error correction codes based on states generated via particle number resolving measurements}

\author{S. B. Korolev, T. Yu. Golubeva} 
\affiliation{St.Petersburg State University, Universitetskaya nab. 7/9, St.Petersburg, 199034, Russia}
\begin{abstract}
We introduce bosonic error correction codes for particle loss and dephasing errors, constructed from states generated by particle number measurements on two-mode Gaussian states. We analyze these states for their suitability in correcting such errors and identify the features most conducive to error mitigation.
\end{abstract}
\maketitle
\section{Introduction}

Quantum error correction has become an essential component of almost every quantum information protocol. Whether quantum information is being transmitted (quantum communication), stored (quantum memory), or processed (quantum computation), it must be protected against unavoidable noise. Only with reliable error protection can one ensure the practical functionality and scalability of quantum protocols.

The protection of quantum information can be achieved using quantum error-correcting codes  \cite{Nielsen_Chuang_2010}. These codes encode quantum information in a way that makes it resilient to specific classes of errors. The underlying principle of any quantum error correction code is redundancy: the information must be encoded in such a way that, even if part of it is lost, the remainder suffices to fully reconstruct the original state. Redundancy can be realized in different ways. A traditional approach encodes a logical qubit into multiple physical qubits \cite{Peres1985,Shor1995,Steane1996}. However, this method requires the creation of large-scale systems of interacting qubits, which remains experimentally challenging.

An alternative approach employs bosonic error correction codes, where logical qubits are encoded in the continuous-variable states of a single quantum oscillator \cite{Leghtas2013,Albert2018,Terhal_2020,Wallraff2004,Terhal2016,Wasilewski2007,Glancy2004}. In this case, redundancy arises from encoding two logical basis states in the infinite-dimensional Hilbert space of the oscillator. A practical advantage of this method is that one logical qubit requires only one oscillator, rather than many ancillary qubits.

Several classes of bosonic codes have been proposed. Gottesman–Kitaev–Preskill (GKP) codes protect against displacement errors \cite{Gottesman2001,Vasconcelos2010} and particle loss errors  \cite{Hastrup2023}. Schrödinger cat codes also provide protection against particle loss errors \cite{Cochrane1999,Bergmann2016,Mirrahimi_2014}. It has been demonstrated that squeezed cat states, squeezed Fock states, and binomial states can simultaneously correct both particle loss and dephasing errors \cite{Schlegel2022,Korolev2024_eror_cor, Binomial_state}.

A key limitation of most existing bosonic codes is that the ideal code states cannot be prepared exactly in an experiment. Only approximate versions of these states are experimentally accessible \cite{Konno2024,Kudra2022,Sychev2017,Ourjoumtsev_cat2007,Polzik2006, Huang2015, Ulanov2016, Gerrits2010, Takahashi2008,Thekkadath2020}, which results in approximate and less effective error correction in practice.

In this work, we seek bosonic codes for correcting both particle loss and dephasing errors within a family of states that can be prepared using particle-number measurements on two-mode Gaussian states. Previous studies have identified squeezed Fock states from this family as suitable for error correction \cite{Korolev2024_eror_cor,Bashmakova2025}. Here, we extend the analysis to the entire family, identifying those states that are optimal for error correction.

This paper is organized as follows. In Sec. II we describe particle loss and dephasing channels and introduce channel fidelity as our performance metric. In Sec. III we define the family of states generated by number measurements on two-mode Gaussian states, present their explicit wave functions, and analyze their properties. In Sec. IV we construct two bosonic codes for loss errors from this family and compare their performance in terms of channel fidelity and experimental feasibility. In Sec. V we propose a code for dephasing errors, evaluate its performance, and discuss its feasibility.

\section{Quantum channels and error correction codes}
\subsection{Channels with particle loss and dephasing errors}
The primary objective of this study is to identify, within a certain family of states to be defined later, those suitable for protecting quantum information against two types of errors: particle loss and dephasing errors. Before turning to the search for such states, we first describe the action of these channels on quantum states.

Let us first consider a quantum channel affected by particle loss errors. The evolution of a quantum state described by the density matrix $\hat{\rho}$ in such a channel is governed by the Lindblad master equation \cite{Lindblad1976}:
\begin{align} \label{eq_lin}
\partial_{t} \hat{\rho}(t)=\kappa_{1} \mathcal{D}[\hat{a}] \hat{\rho}(t),
\end{align}
where $\kappa_{1}$ is the particle loss rate, and the superoperator is defined as
\begin{align}
\mathcal{D}[\hat{J}] \hat{\rho}(t) & =\hat{J} \hat{\rho}(t) \hat{J}^{\dagger}-\frac{\hat{J}^{\dagger} \hat{J} \hat{\rho}(t)+\hat{\rho}(t) \hat{J}^{\dagger} \hat{J}}{2}.
\end{align}
The annihilation and creation operators, $\hat{a}$ and $\hat{a}^{\dagger}$, obey the canonical commutation relation $\left[\hat{a},\hat{a}^{\dagger}\right]=1$. The solution to Eq.~(\ref{eq_lin}) can be expressed in the Kraus decomposition form \cite{KRAUS1971311}:
\begin{align} \label{evol_los}
\hat{\rho}(t)\equiv \mathcal{N}_L \left[\gamma\right]\left(\rho(0)\right)=\sum_{j=0}^{\infty} \hat{K}_{j} \hat{\rho}(0) \hat{K}_{j}^{\dagger},
\end{align}
where the Kraus operators for the loss channel are given by \cite{MUeda_1989,Lee_1994,Chuang_1997}:
\begin{align} \label{Krauss_loos}
    \hat{K}_k=\sqrt{\frac{\gamma^k}{k!}}(1-\gamma)^{\hat{n}/2}\hat{a}^k.
\end{align}
Here $\hat{n}=\hat{a}^{\dagger}\hat{a}$ is the particle number operator, and $\gamma=1-e^{-\kappa_1 t}$ is the dimensionless damping parameter, ranging within $\gamma \in [0,1]$. For subsequent analysis, it will be convenient to rewrite these operators in the Fock basis \cite{Chuang_1997}:
\begin{align} \label{kraus_fock}
\hat{K}_k=\sum _{j=k}^\infty \sqrt{\binom{j}{k}\gamma^k(1-\gamma)^{j-k}}
|j-k\rangle\langle j|.
\end{align}

We now turn to a quantum channel that introduces only dephasing errors. As before, the evolution of quantum states in this channel can be described by Kraus operators as 
\begin{align} \label{kraus}
\hat{\rho}(t)\equiv \mathcal{N}_D\left[\gamma_{\phi}\right] \left(\rho(0)\right)=\sum_{j=0}^{\infty} \hat{D}_{j} \hat{\rho}(0) \hat{D}_{j}^{\dagger},
\end{align}
where the Kraus operators for dephasing errors take the form \cite{Leviant2022quantumcapacity}:
\begin{align}
    \hat{D}_{k}=\sqrt{\frac{\gamma_\phi^k}{k!}}e^{-\frac{\gamma _\varphi}{2}\hat{n}^2}\hat{n}^k.
\end{align}
Here $\gamma\phi$ is the dimensionless dephasing rate. The Fock basis decomposition of these operators is given by
\begin{align} \label{Krauss_deph}
    \hat{D}_{k}=\sum_{j=0}^{\infty}\sqrt{\frac{\gamma_\phi^k }{k!}}e^{-\frac{\gamma _\varphi}{2}j^2}j^k|j\rangle\langle j|.
\end{align}

In realistic quantum channels, both particle-loss and dephasing errors are typically present. Such a combined channel is described by a Lindblad equation with two contributions:
\begin{align}
\partial_{t} \hat{\rho}(t) =\kappa_{1} \mathcal{D}[\hat{a}] \hat{\rho}(t)+\kappa_{2} \mathcal{D}\left[\hat{a}^{\dagger} \hat{a}\right] \hat{\rho}(t), 
\end{align}
Since the Lindblad superoperators for loss and dephasing commute \cite{Leviant2022quantumcapacity}: $\mathcal{D}\left[\hat{a}^{\dagger} \hat{a}\right]\mathcal{D}\left[ \hat{a}\right]=\mathcal{D}\left[ \hat{a}\right]\mathcal{D}\left[\hat{a}^{\dagger} \hat{a}\right]$, the evolution of states in a channel with both loss and dephasing can be expressed as a composition of the two channels:
\begin{align}
   \mathcal{N}_{LD}\left[\gamma,\gamma_\phi\right]= \mathcal{N}_L \left[\gamma\right] \circ  \mathcal{N}_D\left[\gamma_\phi\right]=\mathcal{N}_D \left[\gamma_\phi\right] \circ\mathcal{N}_L\left[\gamma\right].
\end{align}
Since the channels are independent, the effects of particle loss and dephasing errors can be analyzed separately. In this work, we study these two error channels independently and identify two sets of states: one optimized for protecting against particle loss errors, and the other for protecting against dephasing errors. States that can correct both types of errors must lie at the intersection of these sets.

\subsection{Quantum error correction codes}

Having described the quantum channels in which we aim to protect information from errors, we now turn to the encoding procedure. The fundamental principle of quantum error correction is redundant encoding of information, which ensures that the original state can be fully reconstructed even if part of it is lost. In our approach, we employ a bosonic quantum code, where the two logical qubit states $|\overline{0}\rangle$ and $|\overline{1}\rangle$ are encoded in a quantum oscillator. These logical states, referred to as codewords (CW), span the code space $\mathcal{H}_L$, which forms a subspace of the infinite-dimensional Hilbert space.

For the CW $|\overline{0}\rangle$ and $|\overline{1}\rangle_L$ to be applicable for error correction in channels described by a set of Kraus operators $\lbrace\hat{K}_i\rbrace$, the following condition should be satisfied \cite{Knill_1997}:
\begin{align} \label{KL_ideal}
    \hat{P}\hat{K}_i^\dagger \hat{K}_j \hat{P}=\lambda_{ij}\hat{P},
\end{align}
where $\hat{P}=|\overline{0}\rangle\langle \overline{0}|+|\overline{1}\rangle \langle \overline{1}|$ is the projector onto the codeword subspace, and $\lambda_{ij}=\lambda_{ji}^*$ are the elements of a Hermitian matrix independent of the CW. Satisfaction of this condition is both necessary and sufficient for error correction. CW that fulfill this requirement are referred to as perfect codes. Such CW allow complete protection of information in a channel with errors described by the Kraus operators ${\hat{K}_i}$.

Unfortunately, no perfect codes that satisfy condition (\ref{KL_ideal}) exactly have yet been identified for realistic quantum channels. All known bosonic quantum codes are approximate codes, for which condition (\ref{KL_ideal}) only holds approximately \cite{Mandayam2012}:
\begin{align}
    \left|\left|\hat{P}\hat{K}_i^\dagger \hat{K}_j \hat{P}-\lambda_{ij}\hat{P}\right|\right|<\varepsilon,
\end{align}
where $\varepsilon>0$ is an arbitrarily small constant. If the CW satisfy this condition, they can correct errors in the channel only approximately. Moreover, the smaller the value of $\varepsilon$, the better the code performs at error correction.

To quantitatively estimate the benefit of using different CW for error correction in quantum channels, one typically evaluates the channel fidelity. Channel fidelity characterizes how close the initial state transmitted through the channel is to the state recovered after the action of the noisy channel: $ F=F\left(\mathcal{R}\circ \mathcal{N} \left(|\psi\rangle \langle \psi|\right),|\psi \rangle  \right)$, where $\mathcal{R}$ denotes the recovery channel, implementing the information recovery procedure after transmission through the noisy channel $\mathcal{N}$.

In our work, we adopt the transpose channel (TC) \cite{Barnum2002,Ng2010} as the recovery channel, which belongs to the family of channels close to optimal. A distinctive feature of the TC is that once the CW and the error channel $\mathcal{N}$ are specified, the recovery procedure is automatically determined. Unlike in optimal recovery channels \cite{Chuang1997,Kosut2009}, no further fine-tuning of the recovery operation is required. On the one hand, this somewhat restricts the recovery procedure, as errors in recovery may occur. However, these recovery errors can always be quantitatively estimated. On the other hand, a major advantage of the TC is its experimental implementability \cite{Biswas2024}.

The fidelity of a quantum channel in which the error process is followed by recovery via the TC is given by \cite{Zheng2024}:
\begin{align} \label{fid_chan}
F=\frac{1}{4}\left\|\operatorname{Tr}_{L} \sqrt{M}\right\|_{F}^{2}, 
\end{align}
where $\left|\left|\cdot\right|\right|_F$ denotes the Frobenius norm, $\left(\operatorname{Tr}_{L} B\right)_{l, k}=\sum_{\mu} B_{[\mu l],[\mu k]}$ represents the partial trace over the code space indices, and $M$ is the quantum error correction (QEC) matrix. The QEC matrix is defined as
\begin{align} \label{QEC_matrix}
    M_{[\mu l],[\nu k]}:=\left\langle\overline{\mu}\right| \hat{K}_{l}^{\dagger} \hat{K}_{k}\left|\overline{\nu}\right\rangle,
\end{align}
where $|\overline{\mu}\rangle$ and $|\overline{\nu}\rangle$ are orthogonal CW, and$\left\{\hat{K}_{i}\right\}$ are the Kraus operators of the error channel $\mathcal{N}$.

In this work, we use the channel fidelity defined in Eq. (\ref{fid_chan}) to evaluate different states generated within a specific scheme, with the aim of assessing their suitability as codes for correcting particle loss and dephasing errors. We now proceed to describe the set of states under investigation.

\section{Set of States Under Investigation}
\subsection{Generation scheme}
\begin{figure}[H]
    \centering
    \includegraphics[width=0.25\linewidth]{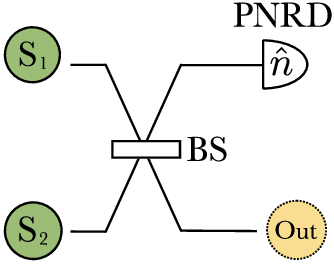}
    \caption{Scheme for generating states evaluated for their suitability as error correcting codes. In the figure: $\text{S}_1$ and $\text{S}_2$ denote squeezed oscillators, BS is a beam splitter, PNRD is a particle number resolving detector, and Out is the output state.}
    \label{fig:scheme}
\end{figure}

As the set of states to be evaluated for their suitability in error correction, we consider those generated in the scheme shown in Fig. \ref{fig:scheme}. 
\noindent In the scheme, two squeezed vacuum states with squeezing parameters $r_1$ and $r_2$ are mixed on a beam splitter with amplitude transmission coefficient $t$. One mode of the resulting two-mode Gaussian state is then measured using a particle number resolving detector (PNRD). For definiteness, we assume that the PNRD registers $m$ particles. As a result, in the second mode, we obtain a non-Gaussian state, which we will use for error correction codes.

As shown in \cite{Korolev_2024}, the states generated by this scheme are finite superpositions of squeezed Fock states. These states have the following state vector:
\begin{align} \label{out_rw}
    \left|\Psi_m^{\text{out}}(r,z)\right\rangle= \begin{cases}
    \hat{\mathcal{S}}\left(r\right)\left(\sum \limits_{k=0}^{m/2}  A_{2k}\left(m,z\right) \left|2k\right\rangle\right), \quad \text{if} \quad m \text{ -- even},\\
    \\
     \hat{\mathcal{S}}\left(r\right)\left(\sum \limits_{k=0}^{(m-1)/2}  A_{2k+1}\left(m,z\right) \left|2k+1\right\rangle\right), \quad \text{if} \quad m \text{ -- odd},
    \end{cases}
\end{align}
where $m$ is the number of particles detected in the scheme, $|k\rangle$ are Fock states, $ \hat{\mathcal{S}}\left(r\right)=\exp\left[\frac{r}{2}\left(\left(\hat{a}^\dag\right)^2-\hat{a}^2\right)\right]$ is the squeezing operator, and the superposition coefficients are given by 
\begin{align} \label{coef_a}
    A_{k}\left(m,z\right)= \frac{\left| z\right| ^{\frac{m-k}{2}}}{ 2^{\frac{m-k}{2}} \left(\frac{m-k}{2}\right)! \sqrt{\,
   _2F_1\left(\frac{1-m}{2},-\frac{m}{2};1;\left| z\right| ^2\right)}}\sqrt{\frac{m!}{k!}},
\end{align}
where $\,
   _2F_1\left(\frac{1-m}{2},-\frac{m}{2};1;\left| z\right| ^2\right)$ is the hypergeometric function.
   
From Eq. (\ref{out_rw}), it follows that the generated state depends on three parameters: $r$, $z$, and $m$. The parameter $r$ determines the squeezing of the Fock states, while $z$ controls the superposition coefficients. These parameters are related to the parameters of the generation scheme as follows \cite{Korolev_2024,Korolev2024_fock,Bashmakova_2023}:
   \begin{align}
       &z=-\frac{ (1-t) \sinh (2 r_1)+t \sinh (2
   r_2)}{2 \sinh^2(r_1-r_2) (1-t) t},\\
  & \cosh (r)=\frac{1}{\sqrt{(1-t \tanh (r_1)-(1-t) \tanh (r_2)) (1+t \tanh
   (r_1)+(1-t) \tanh (r_2))}}.
   \end{align}
   From the above expressions, it is clear that by adjusting the scheme parameters (squeezing degree and transmission coefficient), one can control the characteristics of the output state. The parameters $z$ and $r$ can take any real values. Another parameter characterizing the generated state is the measured particle number $m$. This number determines the number of terms in the expansion of the generated state and also imposes constraints on the superposition coefficients given in Eq. (\ref{coef_a}).
   
A distinctive feature of the considered generation scheme is that the output states are generated probabilistically. Since the generation of the output state (\ref{out_rw}) results from the measurement of a particle number $m$, the probability of generating a particular state is equal to the probability of detecting this particle number. This probability is given by \cite{Korolev_2024}:
\begin{align} \label{prob}
    P_m\left(a,z\right)=\frac{2(a-1)^m}{(a+1)^{m+1}| 1-z| ^{m+\frac{1}{2}}} \sqrt{a(| 1-z| -1)+1} \,
   _2F_1\left(\frac{1-m}{2},-\frac{m}{2};1;| z| ^2\right),
\end{align}
where an additional parameter $a$ is introduced, expressed in terms of the scheme parameters as
\begin{align}
    a=e^{2 r_2} t+e^{2r_1} (1-t).
\end{align}
The parameter $a$ takes real values greater than one ($a>1$). The generated state is thus defined by three parameters: $a$, $z$, and $r$, each expressed in terms of scheme parameters in a linearly independent way. From this linear independence, and from the fact that $a$ appears only in the expression for the probability, it follows that $a$ does not affect the type of generated state but solely determines the probability of its generation. By varying $a$, one can maximize the probability of generating a desired non-Gaussian state \cite{Korolev_2024,Korolev2024_fock}.

\subsection{Characteristics of generated states}
\subsubsection{Parity of states}
Having established which states are generated in the scheme shown in Fig.~\ref{fig:scheme}, we now turn to identifying the characteristics of these states that are relevant for error correction codes.

The first requirement for states to be applicable to particle loss error correction codes is parity. A state is said to possess definite parity if it contains either only even or only odd particle numbers. If, during the course of evolution, such a state changes its initial parity, this signals a particle loss error, which can potentially be corrected.

In the scheme under consideration, the generated states can be decomposed into Fock states as follows:
\begin{align} \label{Fock_0}
  &\left|\Psi_{2m+1}^{\text{out}}(r,z)\right\rangle = \sum_{k=0}^{\infty} \frac{ (-1)^k 
 \sqrt{(2 m+1)! (2 k+1)!}
 \, {}_2F_1\left(-m, -k, \frac{3}{2}, \frac{-1}{\sinh^2 r \left(1 + \frac{z}{\tanh r}\right)}\right) \left(\frac{\tanh r}{2}\right)^{m + k} \left(1 + \frac{z}{\tanh r}\right)^m}{
 m! \, k! \sqrt{{}_2F_1\left(-\frac{1}{2} - m, -m, 1, z^2\right) \cosh^3 r}}\left|2k+1\right\rangle,\\
 &\left|\Psi_{2m}^{\text{out}}(r,z)\right\rangle =\sum _{k=0}^{\infty}  \frac{(-1)^k\sqrt{(2m)! (2k)!}  \, {}_2F_1\left(-m, -k, \frac{1}{2}, \frac{-1}{\sinh^2 r \left(1 + \frac{z}{\tanh r}\right)}\right)\left(\frac{\tanh r}{2}\right)^{m + k} \left(1 + \frac{z}{\tanh r}\right)^m}{m! \, k! \sqrt{{}_2F_1\left(\frac{1}{2} - m, -m, 1, z^2\right) \cosh r}} \left|2k\right\rangle. \label{Fock_1}
\end{align}
We see that the states under consideration are decomposed into an infinite series of Fock states. This arises from the explicit action of the squeezing operator. Furthermore, the expansions reveal that, depending on the parity of the measured particle number (even $2m$ or odd $2m+1$), the generated state contains either an even or an odd number of particles.

This result has a clear physical interpretation. Since the input states in the scheme are squeezed vacuum states, which possess even parity, the generated state must also have definite parity. The parity of the output state is determined by the measurement outcome at the PNRD. Measuring an odd number of particles implies that an odd number of particles has been subtracted from the initial even parity states, leaving the scheme with an odd number of particles. The same reasoning applies in the case of an even number of measured particles. This also accounts for the difference in the decomposition coefficients \cite{Bashmakova_2023}.

\subsubsection{Asymmetry of states on the phase plane}
The next useful characteristic of the states, relevant for correcting dephasing errors, is their asymmetry on the phase plane. As is well known \cite{Arqand2020}, under the action of dephasing errors, a quantum state undergoes a random rotation on the phase plane. If a state exhibits asymmetry on the phase plane, such rotations can be detected, and the errors can therefore be corrected.

The states under consideration are finite superpositions of squeezed Fock states (\ref{out_rw}). By tuning the parameters of the scheme, one can achieve a configuration in which one quadrature of the state is stretched while the other is squeezed. To quantify this asymmetry, we evaluate the quadrature variances of the states obtained upon measuring $m$ particles with the PNRD:
\begin{align}
    &(\Delta x)^2=\langle \hat{x}^2\rangle-\langle \hat{x}\rangle^2=\frac{e^{-2r}}{2}\left(2 m+1  -\frac{m(m-1) z (z-1) \, _2F_1\left(\frac{3-m}{2},1-\frac{m}{2};2;z^2\right)}{ \,
   _2F_1\left(\frac{1-m}{2},-\frac{m}{2};1;z^2\right)}\right),\\
  &(\Delta p)^2=\langle \hat{p}^2\rangle-\langle \hat{p}\rangle^2 =\frac{e^{2 r}}{2} \left(2 m+1-\frac{m (m-1) z (z+1) \, _2F_1\left(\frac{3-m}{2},1-\frac{m}{2};2;z^2\right)}{\,
   _2F_1\left(\frac{1-m}{2},-\frac{m}{2};1;z^2\right)}\right),
\end{align}
where the quadratures are defined in terms of the creation and annihilation operators as follow: $\hat{x}=\left(\hat{a}+\hat{a}^{\dagger}\right)/\sqrt{2}$, $\hat{p}=i\left(\hat{a}^{\dagger}-\hat{a}\right)/\sqrt{2}$. 
An analysis of the resulting expressions shows that the states under consideration will almost always exhibit asymmetry, except in the special case when their parameters satisfy the following condition:
\begin{align}
    e^{4r}=
  \frac{(2m + 1) \, {}_2F_1\left(\frac{1-m}{2}, -\frac{m}{2}, 1, z^2\right) - m (m - 1) z (z - 1) \, {}_2F_1\left(\frac{3-m}{2}, 1 - \frac{m}{2}, 2, z^2\right)}{(2m + 1) \, {}_2F_1\left(\frac{1-m}{2}, -\frac{m}{2}, 1, z^2\right) - m(m - 1) z (z + 1)  \, {}_2F_1\left(\frac{3-m}{2}, 1 - \frac{m}{2}, 2, z^2\right)}.
\end{align}
In this case, the variances are equal, and such states are not suitable for correcting dephasing errors. In all other cases, the states' asymmetry can potentially be exploited for error correction. Moreover, the larger the asymmetry (i.e., the greater the difference between the two variances), the better the states are suited to correcting dephasing errors \cite{Bos_rota_code}. The asymmetry of the states under consideration becomes maximal in the limit $r\rightarrow \infty$, but this limit is unphysical. Therefore, one must search for an optimum by tuning the parameters $r$, $z$, and $m$ while taking experimental feasibility into account. We examine this question in detail below.

\subsubsection{Orthogonality of states}
According to the conditions in Eq. (\ref{KL_ideal}), error correction codes must employ orthogonal CW. For this reason, we need to determine under what circumstances the states under consideration become orthogonal. That is, when the inner product $\left\langle \Psi_{m'}^{\text{out}}(r_1,z_1) \, \middle| \, \Psi_{m}^{\text{out}}(r_2,z_2) \right\rangle$ is zero. The first obvious condition for orthogonality is that  $m$ and $m'$ have different parities. In this case orthogonality is guaranteed regardless of the specific values of $r_1$, $r_2$, $z_1$ and $z_2$. However, such a situation eliminates the advantage of using definite parity to diagnose particle loss errors. If, on the other hand, $m$ and $m'$ have the same parity, then determining the parameter values that satisfy the orthogonality condition requires the use of the following relations:
\begin{multline} \label{orth_1}
   \left\langle \Psi_{2m'+1}^{\text{out}}(r_1,z_1) \, \middle| \, \Psi_{2m+1}^{\text{out}}(r_2,z_2) \right\rangle= \frac{ \left(\frac{\tanh(r_1 - r_2)}{2}\right)^{m' + m}   \left(\frac{z_2}{\tanh(r_1 - r_2)} - 1\right)^m   \left(\frac{z_1}{\tanh(r_1 - r_2)} + 1\right)^{m'}}{m!\, m'!\,}\\
   \times\frac{ \sqrt{(2m+1)! (2m'+1)!}   \,{}_2F_1\left(-m, -m', \frac{3}{2}, \frac{1}{\sinh^{2}(r_1 - r_2) \left(\frac{z_2}{\tanh(r_1 - r_2)} - 1\right) \left(\frac{z_1}{\tanh(r_1 - r_2)} + 1\right)}\right)
    \, }{   \sqrt{  \cosh^3(r_1 - r_2) \,        {}_2F_1\left(-\frac{1}{2} - m, -m, 1, z_2^2\right)         \,  {}_2F_1\left(-\frac{1}{2} - m', -m', 1, z_1^2\right)} },
\end{multline}
\begin{multline} \label{orth_2}
   \left\langle \Psi_{2m'}^{\text{out}}(r_1,z_1) \, \middle| \, \Psi_{2m}^{\text{out}}(r_2,z_2) \right\rangle= \frac{ \left(\frac{\tanh(r_1 - r_2)}{2}\right)^{m' + m}   \left(\frac{z_2}{\tanh(r_1 - r_2)} - 1\right)^m   \left(\frac{z_1}{\tanh(r_1 - r_2)} + 1\right)^{m'}}{m!\, m'!\,} \\
   \times\frac{ \sqrt{(2m)! (2m')!}   \,{}_2F_1\left(-m, -m', \frac{1}{2}, \frac{1}{\sinh^{2}(r_1 - r_2) \left(\frac{z_2}{\tanh(r_1 - r_2)} - 1\right) \left(\frac{z_1}{\tanh(r_1 - r_2)} + 1\right)}\right)
    \, }{   \sqrt{  \cosh(r_1 - r_2) \,        {}_2F_1\left(\frac{1}{2} - m, -m, 1, z_2^2\right)         \,  {}_2F_1\left(\frac{1}{2} - m', -m', 1, z_1^2\right)} }.
\end{multline}
From the above expressions, it follows that for different measurement outcomes $m$ and $m'$, one obtains different parameter relations among $r_1$, $r_2$, $z_1$, and $z_2$ that lead to the orthogonality of the states. Moreover, as the values of $m$ and $m'$  increase, the degree of the hypergeometric functions also increases, resulting in a larger number of parameter relations that yield orthogonal states. In other words, the larger the measured particle number, the greater the number of mutually orthogonal states that can be constructed. Examples of orthogonal states generated within the considered scheme are provided in Appendix \ref{Append_param_state}.

\subsubsection{Mean Particle Number in States}
The final characteristic of the states that we consider is their mean particle number. It turns out that the mean particle number directly affects the probability of particle loss in the channels. From the expression for the Kraus operators of the particle loss channel (\ref{kraus_fock}), it follows that the probability for a quantum state $|j\rangle$ to transition into $|j-k\rangle$  is given by a binomial distribution and depends on the index $j$. This implies that as the mean particle number of a state increases, so does the probability of losing a larger number of particles. In other words, when two states propagate through the same channel (with the same damping parameter $\gamma$), the state with the larger particle number is more likely to lose more particles. For this reason, it is crucial to determine the mean particle number in the states under investigation.

The mean particle number in the state, conditioned on the detection of $m$ particles, is given by
\begin{align}
   \left\langle \Psi_{m}^{\text{out}}(r,z) \, \middle|\hat{n}\middle| \, \Psi_{m}^{\text{out}}(r,z) \right\rangle =\frac{\cosh 2r}{2} \left( 
 2m+1 - \frac{m(m-1)z(z + \tanh 2r) \, {}_2F_1\left(\frac{3-m}{2}, 1-\frac{m}{2}, 2, z^2\right)}
{{}_2F_1\left(\frac{1-m}{2}, -\frac{m}{2}, 1, z^2\right)} 
\right) - \frac{1}{2}.
\end{align}
From this expression, it is straightforward to see that the mean particle number increases with the squeezing parameter $r$, the detection outcome $m$, and the parameter $z$. By tuning these parameters, one can control the mean particle number and thereby select states best suited for correcting particle loss errors. We will return to this point in more detail below.

In summary, the states generated in the scheme depicted in Fig.~\ref{fig:scheme} are suitable for correcting both particle loss and dephasing errors. Moreover, by appropriately adjusting the scheme parameters, the generated states can be tailored to serve as error correction codes optimized for specific types of noise. We now proceed to the task of identifying such states.

\section{Particle loss errors}
\subsection{Channel fidelity for optimal codewords}
Let us begin the search for states suitable for correcting particle loss errors. As already noted, the first step in error correction is to define the CW. We choose the CW to be the states generated by the scheme in Fig.~\ref{fig:scheme} upon detection of the same number of particles $m$:
\begin{align}
    &\left|\overline{0}\right\rangle=\left|\Psi_m^{\text{out}}(r_1,z_1)\right\rangle, \label{1_CW}\\
    &\left|\overline{1}\right\rangle=\left|\Psi_m^{\text{out}}(r_2,z_2)\right\rangle. \label{2_CW}
\end{align}
In this case, the CW have a defined parity, which is essential for correcting particle loss errors. It is worth noting that one could satisfy the parity requirement by considering different detection outcomes $m$ for the two CW. However, this approach is less practical: the two CW would require different physical resources for their generation and would differ significantly in their generation probabilities. For this reason, throughout this work, we restrict our attention to CW obtained by measuring the same number of particles. 

The pair of CW defined in Eqs.~(\ref{1_CW}) and (\ref{2_CW}) depends on five parameters: $r_1$, $r_2$, $z_1$, $z_2$, and $m$. However, in order to satisfy condition (\ref{KL_ideal}), the CW must be orthogonal, i.e.,
\begin{align}
\left\langle \overline{0} \middle|\overline{1}\right\rangle
=\left\langle \Psi_m^{\text{out}}(r_1,z_1)\middle|\Psi_m^{\text{out}}(r_2,z_2)\right\rangle=0.
\end{align}
It should be emphasized that the parameter $m$ can take any value except unity. As demonstrated in Ref.~\cite{Korolev2024_fock}, when $m=1$ the scheme necessarily produces the first squeezed Fock states, which are never orthogonal.

Since we aim to correct particle loss errors, it is important that the mean particle numbers in the CW be identical. Therefore, we impose two additional conditions:
\begin{align}
    \left\langle \overline{0} |\hat{n}|\overline{0}\right\rangle=\left\langle \Psi_m^{\text{out}}(r_1,z_1)|\hat{n}|\Psi_m^{\text{out}}(r_1,z_1)\right\rangle=\left\langle n\right\rangle,\\
   \left\langle \overline{1} |\hat{n}|\overline{1}\right\rangle= \left\langle \Psi_m^{\text{out}}(r_2,z_2)|\hat{n}|\Psi_m^{\text{out}}(r_2,z_2)\right\rangle=\left\langle n\right\rangle.
\end{align}
Thus, out of the five parameters, three remain free. For definiteness, we will use  $r_2$, $\left\langle n\right\rangle$, and $m$ as independent parameters. By varying these three parameters, we will search for CW that are best suited for error correction. We will refer to such CW as optimal CW.

Let us now proceed to evaluate the efficiency of applying the proposed CW to particle loss error correction. As noted earlier, to assess the performance of error correction with CW, one should use the channel fidelity (\ref{fid_chan}), together with the Kraus operators of the particle loss channel (\ref{Krauss_loos}) characterized by the damping parameter $\gamma$. To compute the fidelity, we first need to calculate the QEC matrix (\ref{QEC_matrix}). Since, in general, the number of Kraus operators describing the particle loss channel is infinite, the QEC matrix is infinite-dimensional. Therefore, to enable numerical evaluation of the fidelity, we employ an approximation that truncates the number of Kraus operators describing the channel.

To restrict the number of Kraus operators, we assume that the states under consideration contain a finite number of particles, $N$. In other words, the decomposition of states (\ref{Fock_0}) and (\ref{Fock_1}) are truncated at $N$ rather than extending to infinity. As a result, in the expression for the Kraus operators (\ref{kraus_fock}), the infinite summation limit is replaced by $N$, and hence the number of Kraus operators becomes finite and equal to $N$.

This approximation is physically justified by the fact that, in realistic generated states, the probability of detecting a large number of particles (greater than some $N$) is negligibly small. The value of $N$ is determined by the parameters of the generated state. In our work, the cutoff $N$ is chosen according to two criteria: first, $N$ must be much larger than the mean particle number in the state ($N \gg \langle n\rangle$); second, we fix the calculation accuracy by requiring that the probability of detecting $N+1$ particles in states (\ref{Fock_0}) and (\ref{Fock_1}) be smaller than $10^{-9}$. While such a choice of $N$ may be redundant, it guarantees that the difference between the channel fidelities of the true (infinite-dimensional) and truncated (finite-dimensional) states appears only in the ninth decimal place.

\begin{figure}[H]
    \centering
    \includegraphics[width=0.9\linewidth]{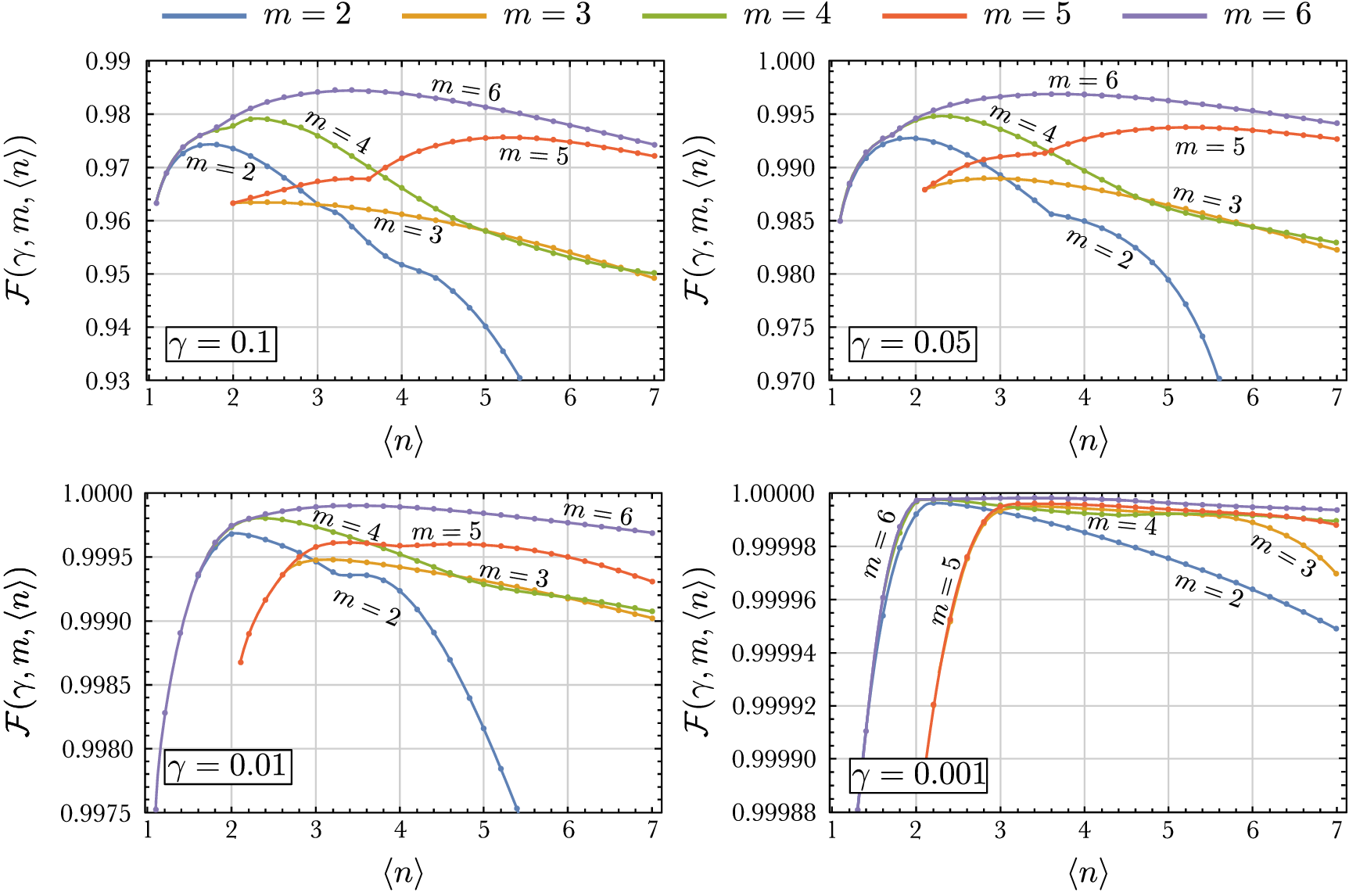}
    \caption{Dependence of the channel fidelity $\mathcal{F}(\gamma, m, \langle n \rangle)$ on the mean particle number  $\langle n \rangle$ in the CW. The plots correspond to channels with different damping parameters $\gamma$. Different colors indicate CW generated by measuring different particle numbers $m$. Points are connected by lines for clarity.}
    \label{fig:fid_states}
\end{figure}

Fig. \ref{fig:fid_states} shows the channel fidelity, maximized over the free parameter $r_2$, as a function of the mean particle number in the CW, $\langle n \rangle$, for different damping parameters $\gamma$ and different measurement outcomes $m$. In other words, we evaluate $\mathcal{F}(\gamma,m,\langle n\rangle)=\max \limits_{r_2} F(r_2,m_,\langle n\rangle)$. Here and throughout, we restrict to $m \leq 6$, since detecting larger particle numbers is unlikely in practice. From Eq. (\ref{prob}) it follows that the probability of measuring $m=7$ particles is below five percent.
\noindent Each point on the plot represents the maximum channel fidelity for a given damping parameter $\gamma$, achievable using the CW (\ref{1_CW}) and (\ref{2_CW}) with a specific value of $m$ and a fixed mean particle number $\langle n \rangle$. The maximization is performed over the free parameter $r_2$.

From Fig. \ref{fig:fid_states}, it is evident that for a fixed value $m$, the quantity $\mathcal{F}(\gamma,m,\langle n\rangle)$ increases with the mean particle number, reaches a maximum, and then decreases. This behavior of the fidelity can be explained by the fact that the particle loss probabilities for a state transmitted through the channel (\ref{Krauss_loos}) depend on both the damping parameter $\gamma$ and the mean particle number of the state. The larger the mean particle number, the higher the probability of losing more than one particle from the state. Losses of more than one particle are poorly corrected by the chosen CW, since they have a certain parity. This leads to a reduction in channel fidelity for large mean particle numbers of the employed states. As for small mean particle numbers, the reduced fidelity in this case is explained by the fact that the loss of even a single particle induces irreversible changes to the states that are difficult to recover. For example, if a state contains on average one particle, the loss of that particle essentially destroys the state. The maximal fidelity therefore lies between these two extremes: when the state contains enough particles that information is not entirely lost upon the removal of one particle, but not so many particles that the probability of losing multiple particles becomes significant.

\begin{figure}[H]
    \centering
    \includegraphics[width=0.55\linewidth]{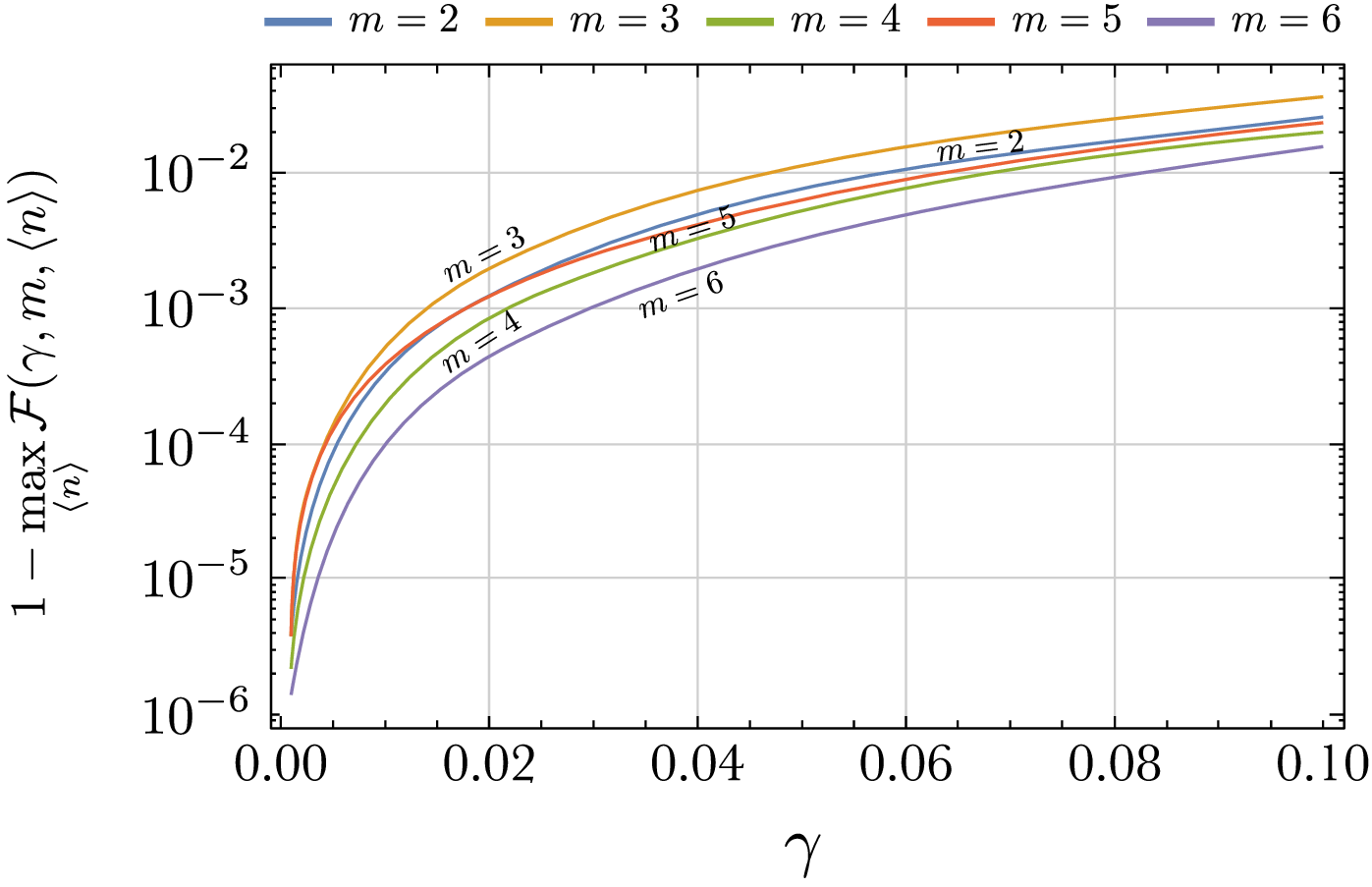}
    \caption{Dependence of the minimal channel infidelity $1-\max\limits_{\langle n\rangle} \mathcal{F}(\gamma,m,\langle n\rangle)$ on the damping parameter $\gamma$. Different colors correspond to CW generated by measuring different particle numbers $m$.}
    \label{fig:opt_fid_max}
\end{figure}

Having established that the channel fidelity obtained with the CW (\ref{1_CW}) and (\ref{2_CW}) exhibits a maximum as a function of the mean particle number, we now turn to the analysis of the maximal fidelity achievable in a channel with damping parameter $\gamma$. For clarity, we plot the channel infidelity minimized over both $r_2$ and $\langle n\rangle$, given by $1-\max \limits_{\langle n\rangle} \mathcal{F}(\gamma,m,\langle n\rangle)$, as a function of the damping parameter $\gamma$. This plot is shown in Fig.~\ref{fig:opt_fid_max}.
\noindent From Fig. \ref{fig:opt_fid_max} it can be seen that for damping parameters $\gamma \leq 0.06$, nearly all CW (except for the case $m=3$) yield channel infidelities below $0.01$. At $\gamma=0.1$, the minimal infidelity is approximately $0.015$, achieved for the CW with $m=6$. To reach smaller values of the infidelity, one needs to employ CW generated by measuring larger particle numbers.

Furthermore, Fig. \ref{fig:opt_fid_max} demonstrates that, in general, even-$m$ CW are more effective for correcting particle loss errors than odd-$m$ ones. Specifically, for all damping parameters $\gamma$, the performance with $m=2$ surpasses that with $m=3$, the case $m=4$ outperforms $m=5$, and so on.

This behavior can be attributed to two factors. First, as $m$ increases, the Fock-state distribution of the optimal CW becomes progressively more uniform (see Appendix \ref{Append_A} for details). As is well known \cite{Bos_rota_code}, the most suitable CW for particle loss error correction are those with uniform weight across all particle numbers, i.e., with decomposition coefficients of equal modulus. However, such states are idealized and not physically realizable. In experimentally generated CW, the distribution cannot be globally uniform, but becomes approximately uniform within a limited range of particle numbers.

Second, the maximum channel fidelity for even-$m$ and odd-$m$ CW occurs at different mean particle numbers. As seen in Fig. \ref{fig:fid_states}, the mean particle number for CW with $m=3$ exceeds that of the $m=2$ and $m=4$ cases, while for $m=5$ it is larger than for $m=6$. As noted earlier, larger mean particle numbers generally degrade the ability of a state to correct particle-loss errors. On the other hand, increasing the mean particle number allows the Fock-state distribution to become more uniform. Thus, optimizing the CW amounts to finding a compromise between these two competing effects.

\subsection{Channel fidelity for identical code words rotated on the phase plane}
In the previous section, we identified the CW that are best suited for correcting particle loss errors. These CW are generated within the scheme shown in Fig. \ref{fig:scheme}, with the same number of detected particles $m$, but with different beam splitter parameters and input squeezing. An example of the parameters required to generate two such optimal CW is given in Appendix \ref{Append_primeri}. Since generating a CW requires two distinct parameter sets, the practical use of these states is significantly complicated.

To simplify the generation procedure for the CW (\ref{1_CW}) and (\ref{2_CW}), we seek pairs of CW that do not require substantial reconfiguration of the scheme. It turns out that one can employ CW that differ only by a $\pi/2$ rotation on the phase plane, while still remaining orthogonal. In this case, the two generation schemes differ solely by the inclusion of a phase-shifting element that implements the rotation. This simplification, however, comes at the cost of reduced channel fidelity. The primary question addressed in this subsection is the extent to which this simplification of the generation scheme affects the achievable fidelity.

For the CW (\ref{1_CW}) and (\ref{2_CW}) to satisfy this simplified generation condition, their parameters must be related by $r_1=-r_2$ and $z_1=-z_2$.
\begin{align}
    &\left|\overline{0}\right\rangle=\left|\Psi_m^{\text{out}}(-r_2,-z_2)\right\rangle,\label{3_CW}\\
    &\left|\overline{1}\right\rangle=\left|\Psi_m^{\text{out}}(r_2,z_2)\right\rangle. \label{4_CW}
\end{align}
We will refer to these states as rotated CW.

In addition, we still require orthogonality of the CW:
\begin{align}
  \left\langle \overline{0} |\overline{1}\right\rangle=\left\langle \Psi_m^{\text{out}}(-r_2,-z_2)|\Psi_m^{\text{out}}(r_2,z_2)\right\rangle=0.
    \end{align}
 At the same time, it is not necessary to impose an additional constraint on the equality of the mean particle number. This condition is automatically satisfied since the CW are identical states rotated with respect to each other on the phase plane. As a result, these CW retain only a single free parameter, namely the mean particle number $\langle {n}\rangle$. By optimizing over $\langle {n}\rangle$, we can determine the maximum channel fidelity achievable with these CW and compare it with the maximum fidelity obtained in the previous section. 
 
\begin{figure}[H]
    \centering
    \includegraphics[width=0.55\linewidth]{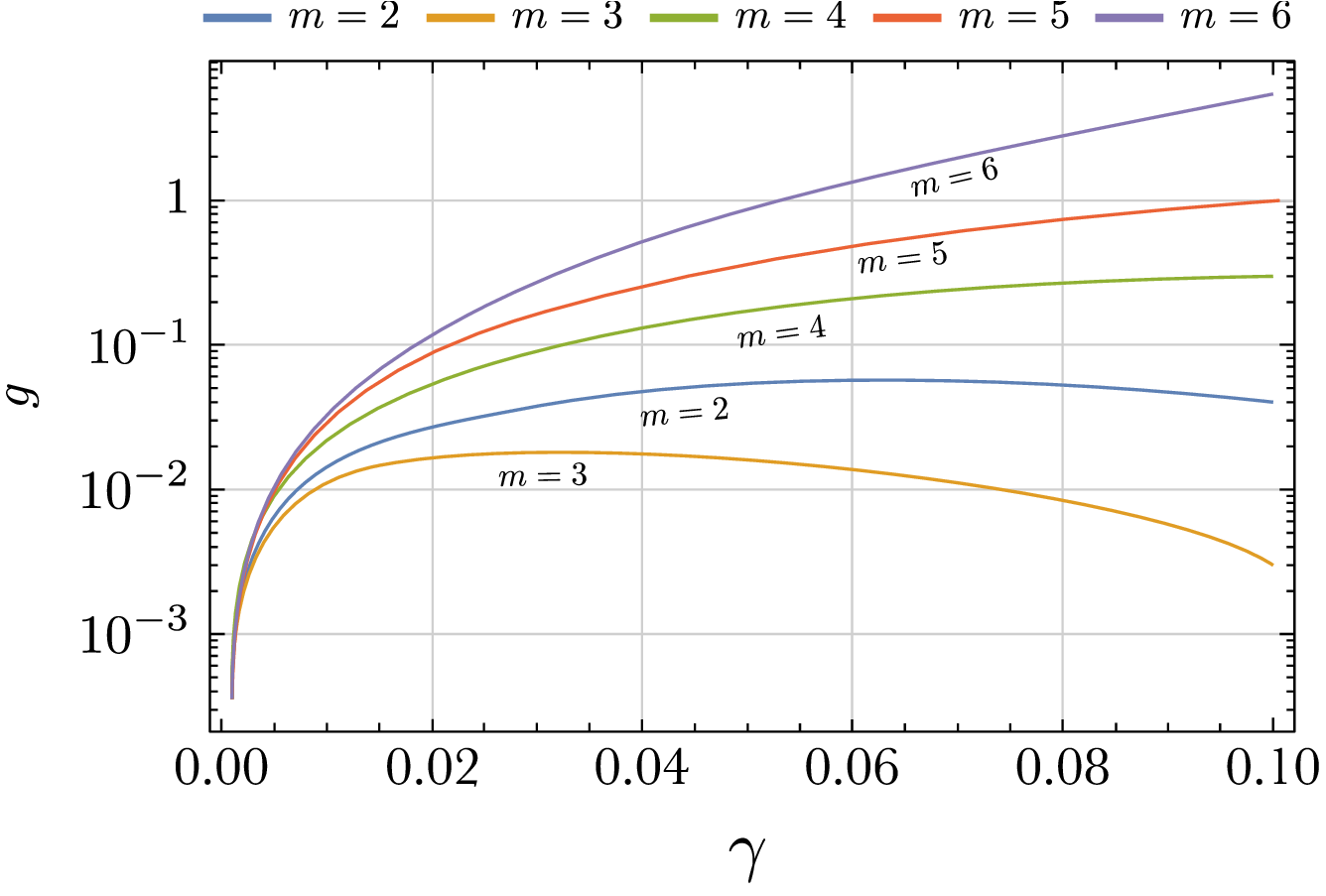}
    \caption{Relative gain in the maximum channel fidelity achieved with the optimal CW compared to that obtained with the rotated CW, as a function of the damping parameter $\gamma$. Different colors indicate CW generated by measuring different particle numbers $m$.}
    \label{fig:gain_opt_nonopt}
\end{figure}

Fig. \ref{fig:gain_opt_nonopt} shows the relative gain in the maximum channel fidelity achieved with the optimal CW compared to that obtained with the rotated CW ( (\ref{3_CW}) and (\ref{4_CW})), for different damping parameter $\gamma$. The gain is defined as $g=\left(\frac{F_{opt}}{F_{rot}}-1\right) \times 100 \% $.
\noindent Fig. \ref{fig:gain_opt_nonopt} shows that the importance of optimization grows with the damping parameter and the number of detected particles $m$. However, within experimentally relevant values of $m$, the difference between the two types of CW  remains modest, with a maximum relative gain $g$ of about  $5\, \%$ at $m=6$ and $\gamma=0.1$. Thus, under realistic experimental constraints on the number of detected particles, the strategy of using CW that are easier to implement experimentally performs only slightly worse than the one based on optimal CW.

\subsection{Comparison of codewords from the perspective of experimental implementation}
Having compared the two types of CW in terms of channel fidelity, we now turn to their comparison from the perspective of experimental implementation. We focus on CW obtained by detecting an even number of particles ($m=2$, $m=4$, and $m=6$), as they exhibit an advantage in correcting particle loss errors.

In practice, two key characteristics determine the feasibility of generating CW: the probability of their generation and the required physical resources. Since the construction relies on squeezed vacuum states, the primary resource is the squeezing of the input oscillators, which is known to be severely limited \cite{Vahlbruch_2016}. The current experimental record stands at 15 dB of squeezing, making it essential that the required squeezing be within experimentally achievable limits.

\begin{figure}[H]
    \centering
    \includegraphics[width=1\linewidth]{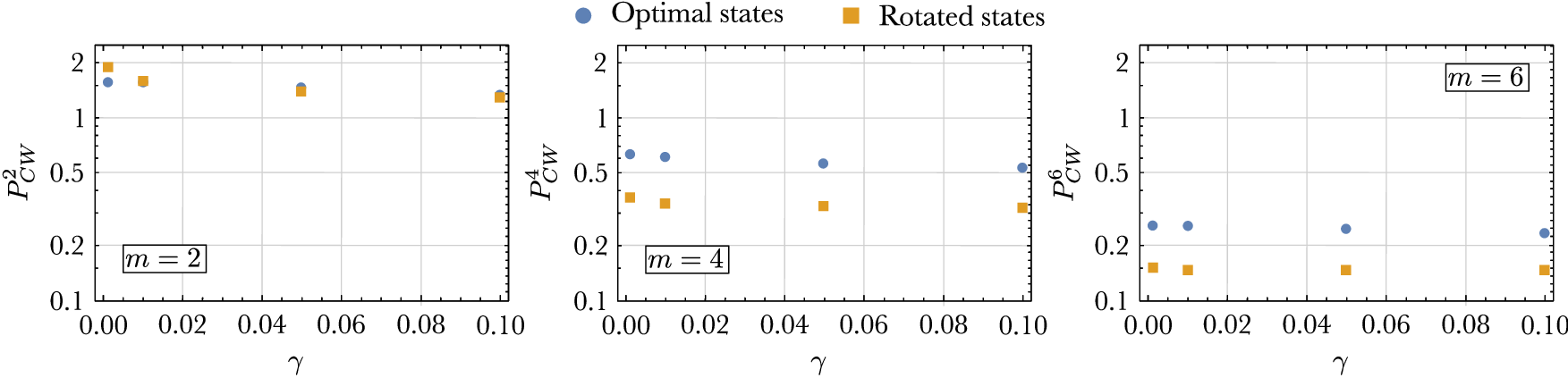}
    \caption{Joint probability of generating two CW as a function of the damping parameter $\gamma$. The plots correspond to CW obtained by detecting $m=2$ (left), $m=4$ (center), and $m=6$ (right) particles. Two cases are shown: optimal CW (blue circle) and rotated CW (yellow square).}
    \label{fig:prod_prob}
\end{figure}

We begin our comparison with the generation probabilities. Note that the two optimal CW may be generated with different probabilities; thus, we define the probability of generating a CW as the product of the probabilities of generating each codeword individually. The probability of generating each codeword is given by Eq. (\ref{prob}). As discussed earlier, this expression contains a free parameter $a$, which can be optimized to maximize the probability. By maximizing over $a$ and substituting the parameter $z$ obtained from the channel fidelity optimization, we obtain the probability of generating a pair of CW best suited for correcting errors at a given damping parameter $\gamma$:  $P_{CW}^m=P_m(a_1,z_1)P_{m}(a_2,z_2)\times 100 \%$. The joint probability $P_{CW}^m$ as a function of $\gamma$ is shown in Fig. \ref{fig:prod_prob} for both the optimal and rotated CW.
\noindent From the figure, one can see that the generation probability decreases with increasing number of detected particles $m$. For $m=2$, the joint probability of generating the CW reaches a few percent, whereas for $m=6$ it drops to fractions of a percent. Thus, increasing the number of detected particles enhances the channel fidelity but reduces the probability of generating the CW. Moreover, the figure indicates that optimal CW generally exhibit higher probabilities than rotated CW. However, in absolute terms, this advantage remains rather modest.

\begin{figure}[H]
    \centering
    \includegraphics[width=1\linewidth]{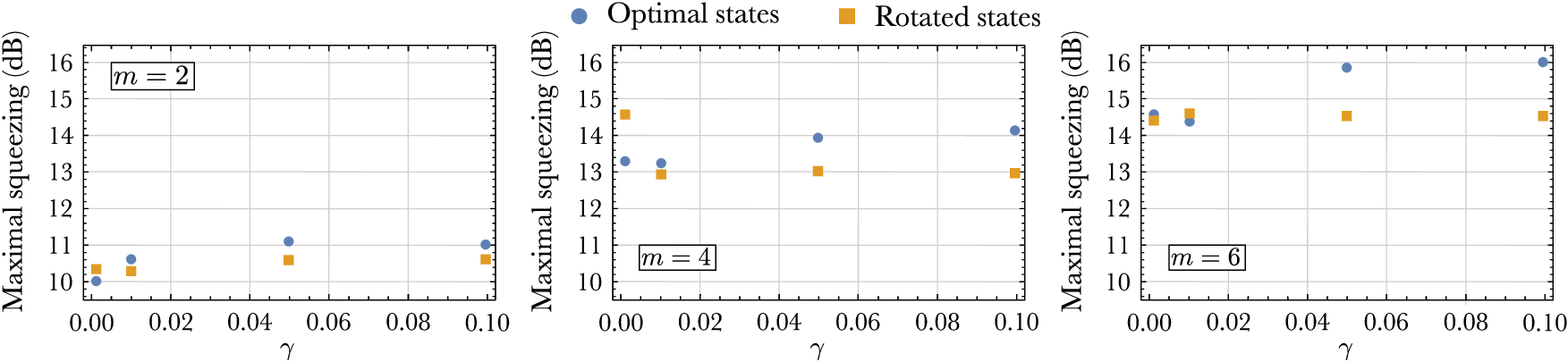}
    \caption{Maximum squeezing of the oscillator required for generating the CW as a function of the damping parameter $\gamma$. The plots correspond to CW obtained by detecting $m=2$ (left), $m=4$ (center), and $m=6$ (right) particles. Two cases are shown: optimal CW (blue circle) and rotated CW (yellow square).}
    \label{fig:Max_sq_pl}
\end{figure}

We now turn to evaluating CW generation in terms of the required squeezing of the oscillators. Here, we consider both optimal and rotated CW that provide the best error correction in a channel with damping parameter $\gamma$ and are generated with the maximum probability $P_{CW}$. Fig.~\ref{fig:Max_sq_pl} shows the maximum oscillator squeezing required for CW generation as a function of $\gamma$. Each point in the plot corresponds to the largest squeezing degree among the four oscillator resources needed to generate a CW.
\noindent The plot shows that generating optimal CW requires oscillators with larger squeezing across almost the entire range of $\gamma$. The situation changes only at a very small damping parameter. It is also evident that increasing the number of detected particles $m$ necessitates higher maximum squeezing. As a result, optimal CW with $m=6$ require oscillators with squeezing beyond experimentally achievable values. Such issues do not arise for rotated CW.

Thus, one can conclude that rotated CW are inferior to optimal CW in terms of generation probability but superior in terms of the maximum squeezing of the required oscillators. Importantly, the loss in probability amounts to only tenths of a percent, while the gain in squeezing averages about 1 dB. Given the experimental challenges of generating highly squeezed states, the advantage in squeezing outweighs the drawback in probability. Therefore, from the perspective of state generation, rotated CW are significantly more practical than optimal CW.

\section{Dephasing errors}
\subsection{Channel fidelity for optimal codewords}
Let us now turn to the procedure of dephasing error correction. For this purpose, we employ the CW defined in Eqs. (\ref{1_CW}) and (\ref{2_CW}), choosing their parameters such that the states are orthogonal, have the same mean particle number, and maximize the channel fidelity under dephasing errors. As in the case of particle loss correction, two parameters remain for optimization; for definiteness, we take $r_2$ and $\langle n\rangle$.

\begin{figure} [H]
    \centering
    \includegraphics[width=1\linewidth]{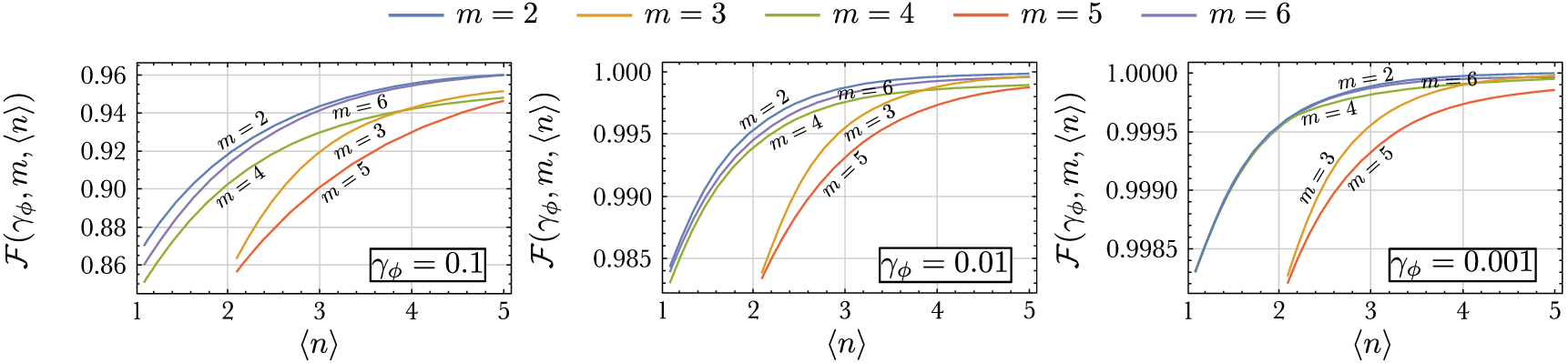}
    \caption{Channel fidelity under dephasing error correction using the optimal CW. The figure shows $\mathcal{F}(\gamma_{\phi},m,\langle n\rangle)$ for different dephasing rates $\gamma_\phi$ and for CW corresponding to different measured particle numbers $m$.}
    \label{fig:deph_fid}
\end{figure}

Fig. \ref{fig:deph_fid} shows the channel fidelity under dephasing, maximized over $r_2$, as a function of the mean particle number $\langle n\rangle$ for different measured particle numbers $m$ and dephasing rates $\gamma_\phi$.
\noindent From the figure, it is evident that, unlike in the case of particle loss, the channel fidelity exhibits no maximum. Instead, it increases monotonically with the average particle number $\langle n\rangle$ for any dephasing rate and any $m$. Regardless of $m$, all fidelities eventually approach unity. However, since even-$m$ CW start from a lower average particle number, they reach near-unit fidelity more rapidly.

An important result is that the optimal CW depend on the average particle number, but not on the dephasing rate in the channel. This distinguishes dephasing error correction from particle loss error correction. For any $\gamma_{\phi}$, the optimal CW coincide whenever their average particle numbers match. This is practically advantageous, as it eliminates the need to adapt CW to the channel: one always selects the optimal CW with the largest experimentally accessible average particle number to achieve the best performance under dephasing error.

It is also worth noting that the CW found through optimization are identical states rotated by $\pi/2$ on the phase plane. In other words, the CW optimal for dephasing error correction coincide with the rotated CW defined in Eqs. (\ref{3_CW}) and (\ref{4_CW}). This is consistent with the intuition that dephasing correction requires maximizing the distance between states on the phase plane \cite{Bos_rota_code}. Thus, CW must be well distinguishable under a phase rotation. Maximum distinguishability is achieved in the limit of infinite squeezing, or equivalently, infinite average particle number, where the structure of the CW disappears and the states collapse into orthogonal lines on the phase plane. This fully agrees with our findings.

For finite, experimentally achievable squeezing, Fig. \ref{fig:deph_fid} shows that the best performance for any dephasing rate $\gamma_\phi$ and average particle number $\langle n\rangle$ is obtained with CW generated by measuring two particles ($m=2$). These CW have the simplest structure and exhibit minimal overlap under phase rotations. For this reason, they also approach unity fidelity more rapidly as squeezing increases.

In summary, the most effective correction of dephasing errors is achieved with the rotated CW generated by measuring two particles ($m=2$). Moreover, the larger the squeezing of such states, the better suited they are for dephasing error correction.

\subsection{Experimental characteristics of the optimal CW}
Let us now evaluate the optimal CW obtained by measuring two particles ($m=2$) from the perspective of experimental feasibility. As before, we will assess both the joint probability of the CW generation, $P_{CW}^2$, and the squeezing required for their generation.

\begin{figure}[H]
    \centering
    \includegraphics[width=0.8\linewidth]{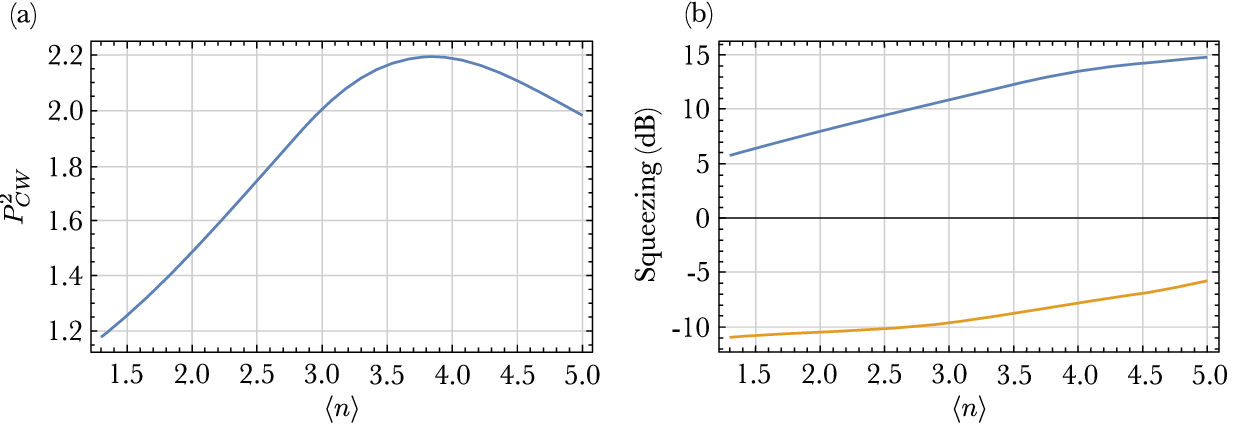}
    \caption{Plots of the joint probability of CW generation (frame (a)) and the squeezing of the required oscillators (frame (b)) as functions of the average particle number in the CW. Both plots correspond to CW obtained by measuring $m=2$ particles. In frame (b), different colors indicate the squeezing of the two oscillators required to generate the CW.}
    \label{fig:res_n2_deph}
\end{figure}

Figure \ref{fig:res_n2_deph}(a) shows the maximum joint probability $P_{CW}^2$ for generating CW best suited for dephasing error correction as a function of the average particle number in the CW, $\langle n\rangle$.
\noindent The figure demonstrates that the probability of generating two CW ranges from $1.2\, \%$ to $2.2\, \%$. This probability reaches its maximum at $\langle n\rangle = \tfrac{1}{2}(5\sqrt{3}-1)$, where the optimal CW are orthogonal squeezed Fock states. As demonstrated in \cite{Korolev_2024}, squeezed Fock states are generated with the highest probability in the considered scheme.

Fig. \ref{fig:res_n2_deph} (b) shows the squeezing (in dB) of the input oscillators required to generate CW optimally suited for dephasing error correction and produced with maximal probability, as a function of the average particle number $\langle n\rangle$. Since the CW differ only by a phase plane rotation, identical schemes are used for their generation. Across the entire range, the required squeezing remains below the experimentally accessible limit of 15 dB. For one oscillator, the squeezing approaches this limit at $\langle n\rangle=5$, indicating that further increases in $\langle n\rangle$ exceed current experimental capabilities.

It should also be noted that CW with $m=6$ exhibit only slightly lower channel fidelity compared to those with $m=2$ (see Fig. \ref{fig:deph_fid}). However, such CW are less favorable from an experimental perspective: their generation probability is an order of magnitude lower, and they require significantly greater resources. Additional details are provided in Appendix \ref{Append_m_6}.




\section{Conclusion}
We have analyzed the suitability of states generated in a scheme based on particle number measurements on a two-mode Gaussian state for use in quantum error correction codes against particle loss and dephasing.

The properties of the states obtained in this scheme were identified and studied. It has been demonstrated that by varying the generation parameters, one can produce quantum states that are potentially suitable for correcting both particle loss and dephasing errors. Based on these findings, we have proposed encodings that protect quantum information from these two types of errors.

For particle loss error correction, states generated by measuring an even number of particles were found to be preferable. Moreover, the larger the number of measured particles, the more advantageous the states become for error correction. Furthermore, it has been demonstrated that the channel fidelity for these states exhibits a maximum that depends on both the average particle number and the damping parameter. Thus, we have found that for any damping parameter, one can select CW that optimally correct the error.

We also considered alternative rotated codewords (CW), differing only by a relative $\pi/2$ phase shift between the states. Analysis of the channel fidelity revealed that these CW perform slightly worse than the optimal ones. However, for CW that are experimentally feasible, the fidelity loss is below one percent.

A comparison of the two types of CW from the perspective of practical implementation showed that optimal CW have higher generation probabilities than rotated ones. This advantage, however, comes at the cost of requiring stronger squeezing of the oscillators used for their generation. Since the degree of experimentally achievable squeezing is severely limited, optimal CW are considerably more challenging to realize in practice.

We further investigated correction for dephasing errors. As in the particle loss case, we identified CW that optimally correct phase errors. These optimal states turned out to be related by a relative $\pi/2$ phase shift. Their performance improves with increasing average particle number.

Among all states analyzed, those generated by measuring two particles were found to be the most effective. They not only provide the highest generation probability but also require squeezing levels that remain within current experimental capabilities.

The fact that the same family of states can be used for correcting both particle loss and dephasing errors implies that they are applicable in channels where both error types occur simultaneously. The optimal CW in such channels differ slightly from those identified in the individual cases: correcting particle loss requires states with a certain average particle number tuned to the damping parameter, while correcting dephasing favors states with as large an average particle number as possible. Consequently, no single state perfectly corrects both error types simultaneously. In realistic channels, one must therefore seek a compromise CW based on the relative error rates. In practice, this means analyzing the joint channel with specified error parameters and identifying the optimal CW accordingly.

The best candidates for such a compromise CW are found among the family of rotated CW states generated by measuring two particles. These rotated CW thus represent the most promising candidates for practical implementation, as they simultaneously mitigate both loss and dephasing errors while offering the highest generation probabilities with minimal resource requirements.

\section*{Funding}
This research was supported by the Theoretical Physics and Mathematics Advancement Foundation "BASIS" (Grant No. 24-1-3-14-1).

\section*{Disclosures}
The authors declare no conflicts of interest.

\section*{Data availability} Data underlying the results presented in this paper are not publicly available at this time, but may be obtained from the authors upon reasonable request.

\bibliography{bibliography}
\appendix
\section{Advantage of even states over odd ones in particle loss error correction} \label{Append_A}
\begin{figure}[H]
    \centering
    \includegraphics[width=0.9\linewidth]{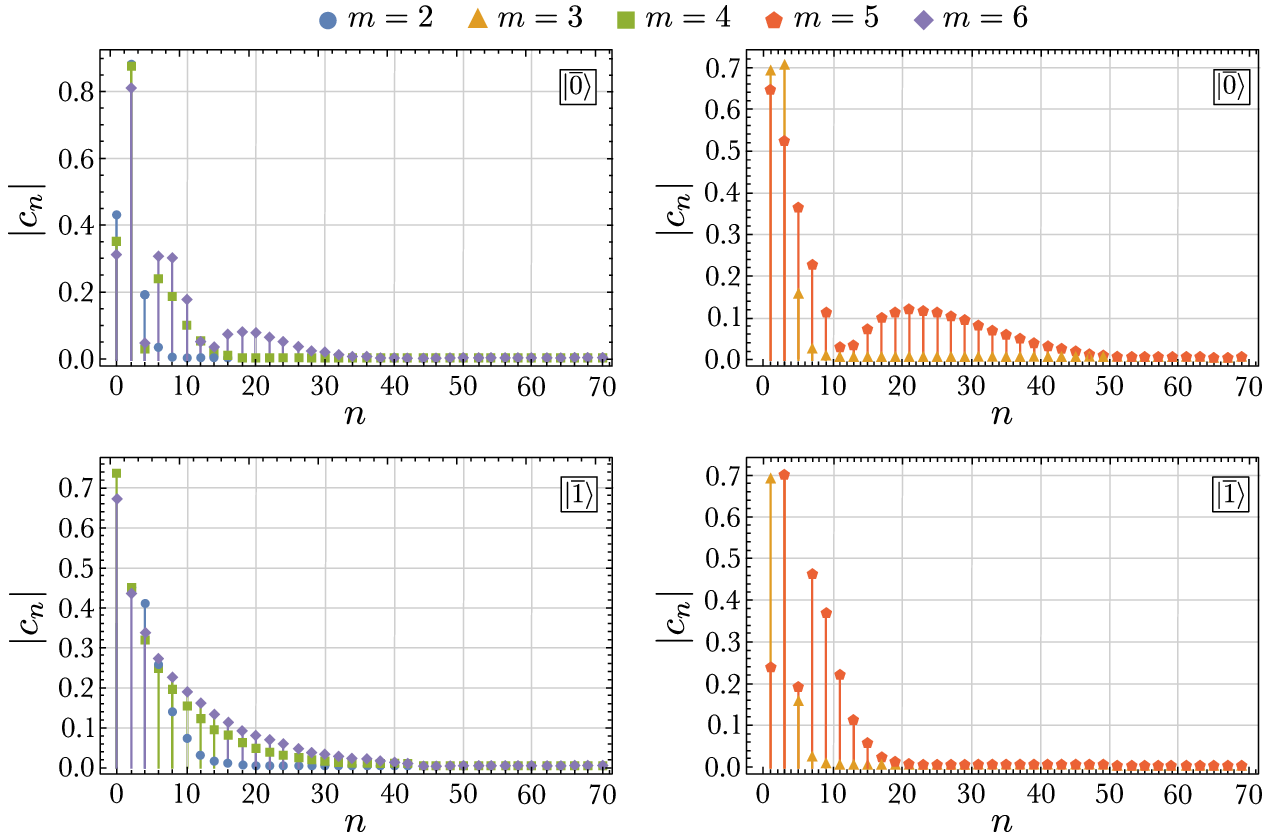}
    \caption{Distribution of the absolute values of the Fock state decomposition coefficients for the CW optimal for correcting particle loss errors with the damping parameter of $\gamma = 0.1$. The left frames show CW generated by measuring an even number of particles ($2$, $4$, and $6$), while the right ones show those generated by measuring an odd number of particles ($3$ and $5$).}
    \label{fig:Fock_distr}
\end{figure}

To understand the origin of the dependence of channel fidelity on the number of measured particles $m$, let us plot the absolute values of the Fock state decomposition coefficients (\ref{Fock_0}) and (\ref{Fock_1}) for the CW that are best suited for correcting particle loss errors (i.e., those yielding the maximum channel fidelity at a given $\gamma$). Fig. \ref{fig:Fock_distr} shows this plot for a damping parameter of $\gamma=0.1$.
\noindent From the figure, it can be seen that the decomposition of the optimal CW behave similarly for both even and odd numbers of measured particles $m$. At the same time, as $m$ increases, the Fock state distribution of the optimal CW becomes increasingly uniform. The same trend is observed for CW optimized for other damping parameters $\gamma$.

\begin{figure}[H]
    \centering
    \includegraphics[width=1.0\linewidth]{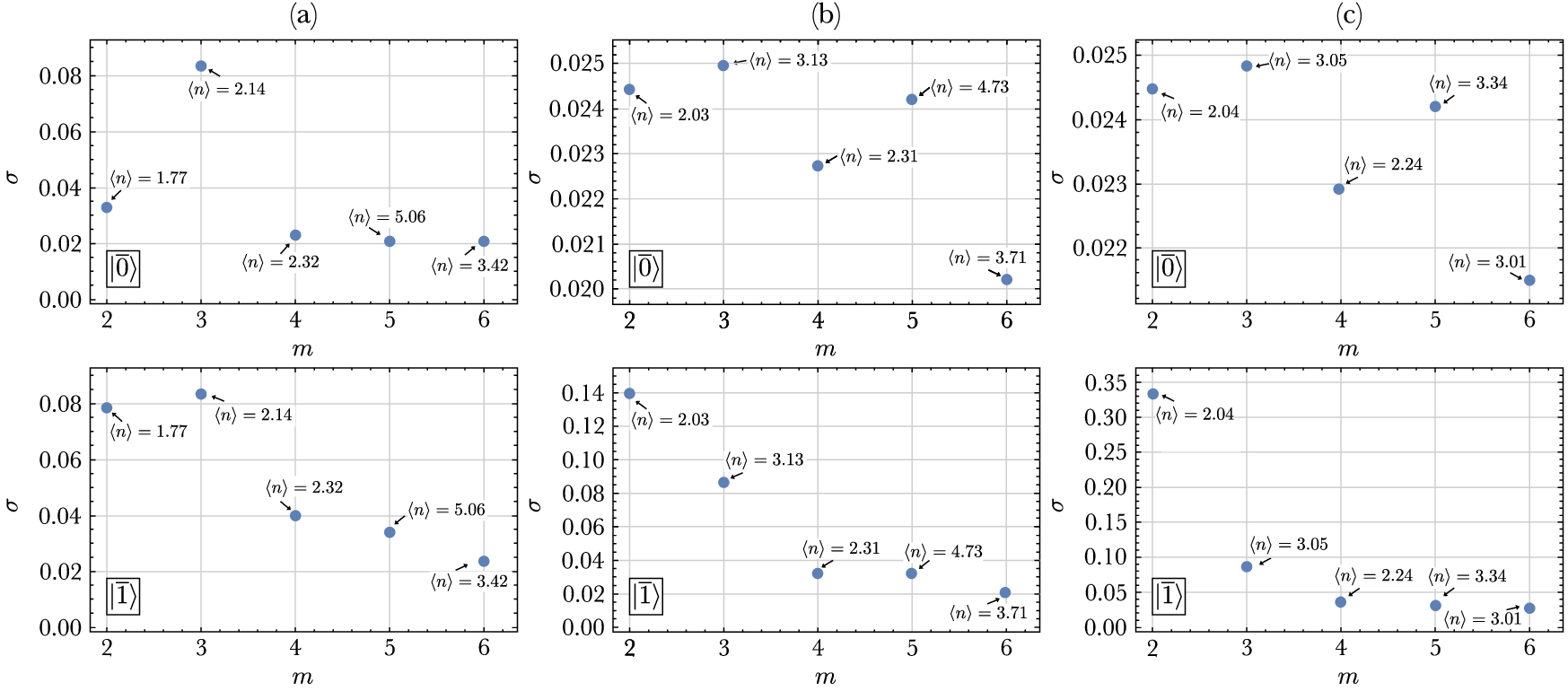}
    \caption{Variances $\sigma$ of the absolute values of the Fock state decomposition coefficients of the optimal CW as a function of the number of measured particles $m$. Each point on the graph is additionally labeled with the corresponding average particle number of the optimal CW. The figure shows three cases: (a) $\gamma=0.1$, (b) $\gamma=0.01$, and (c) $\gamma=0.001$.}
    \label{fig:disp}
\end{figure}

To quantify the degree of “uniformity” of the distributions, let us calculate the variance $\sigma$ of the set of absolute values of the decomposition coefficients of the optimal CW. As the distribution approaches uniformity, its variance tends to zero. Fig. \ref{fig:disp} shows the variance of the distribution of the absolute values of nonzero coefficients (greater than $10^{-9}$) in the Fock state decomposition as a function of the number of measured particles $m$ for different damping parameters $\gamma$.
From the figure, it can be seen that for CW optimal for error correction at different damping parameters $\gamma$, the variance of the distributions of the absolute values of the Fock state decomposition coefficients (\ref{Fock_0}) and (\ref{Fock_1}) decreases with increasing $m$, both for even and odd $m$. However, the relationship between even and odd values of $m$ is not straightforward. We observe that for one of the two CW, the case $m=3$ always performs worse than all other cases, while the case $m=5$ performs worse than $m=4$. This behavior is related to the differences in the decomposition coefficients for even and odd Fock states (\ref{Fock_0}) and (\ref{Fock_1}).

Another important point is that the maximum channel fidelity for even and odd CW is achieved at different average particle numbers. In Fig. \ref{fig:disp}, each point in the variance plot is labeled with the average particle number of the state obtained when measuring a particular $m$. The figure shows that the average particle number in optimal odd CW with $m=3$ and $m=5$ is always larger than in the CW with $m=2$ and $m=6$, respectively.
\section{Parameters of orthogonal states generated in the scheme} \label{Append_param_state}
\subsection{The case of measuring two particles}

\begin{table}[H]
\centering
\begin{tabular}{|l|lll|ll|}
\hline
\multirow{2}{*}{m=2} & \multicolumn{3}{l|}{Experimental parameters}                             & \multicolumn{2}{l|}{State parameters} \\ \cline{2-6} 
                     & \multicolumn{1}{l|}{$\text{S}_1 \, \text{(dB)}$} & \multicolumn{1}{l|}{$\text{S}_2 \, \text{(dB)} $} & t    & \multicolumn{1}{l|}{r}       & z      \\ \hline
$|\overline{0}\rangle$                 & \multicolumn{1}{l|}{2.00}       & \multicolumn{1}{l|}{-3.00}      & 0.08 & \multicolumn{1}{l|}{-0.30}    & -7.14  \\ \hline
$|\overline{1}\rangle$               & \multicolumn{1}{l|}{3.98}       & \multicolumn{1}{l|}{-4.47}      & 0.47 & \multicolumn{1}{l|}{-0.05}   & 0.02   \\ \hline
\multicolumn{6}{|l|}{} \\ \hline
$|\overline{0}\rangle$             & \multicolumn{1}{l|}{4.00}       & \multicolumn{1}{l|}{-5.00}      & 0.23 & \multicolumn{1}{l|}{-0.31}   & -0.89  \\ \hline
$|\overline{1}\rangle$                & \multicolumn{1}{l|}{8.74}       & \multicolumn{1}{l|}{-3.81}      & 0.78 & \multicolumn{1}{l|}{0.55}    & -0.04  \\ \hline
\end{tabular}
\caption{Parameters of two pairs of orthogonal states obtained by measuring $m=2$ particles. The table lists the parameters of the generated state ($r$ and $z$), as well as the experimental parameters of the generation scheme (the squeezing of the input oscillators $\text{S}_1$, $\text{S}_2$ in dB, and the beamsplitter amplitude transmission coefficient $t$).}
\end{table}

\subsection{The case of measuring three particles}

\begin{table}[H]
\centering
\begin{tabular}{|l|lll|ll|}
\hline
\multirow{2}{*}{m=3} & \multicolumn{3}{l|}{Experimental parameters}                             & \multicolumn{2}{l|}{State parameters} \\ \cline{2-6} 
                     & \multicolumn{1}{l|}{$\text{S}_1 \, \text{(dB)}$} & \multicolumn{1}{l|}{$\text{S}_2\, \text{(dB)} $} & t    & \multicolumn{1}{l|}{r}       & z      \\ \hline
$|\overline{0}\rangle$                 & \multicolumn{1}{l|}{2.00}       & \multicolumn{1}{l|}{-3.00}      & 0.08 & \multicolumn{1}{l|}{-0.30}    & -7.14  \\ \hline
$|\overline{1}\rangle$                 & \multicolumn{1}{l|}{3.16}       & \multicolumn{1}{l|}{-6.55}      & 0.31 & \multicolumn{1}{l|}{-0.35}   & 0.14   \\ \hline
\multicolumn{6}{|l|}{} \\ \hline
$|\overline{0}\rangle$                 & \multicolumn{1}{l|}{4.00}       & \multicolumn{1}{l|}{-5.00}      & 0.23 & \multicolumn{1}{l|}{-0.31}   & -0.89  \\ \hline
$|\overline{1}\rangle$               & \multicolumn{1}{l|}{8.071}      & \multicolumn{1}{l|}{-2.09}      & 0.76 & \multicolumn{1}{l|}{0.55}    & -0.48  \\ \hline
\end{tabular}
\caption{Parameters of two pairs of orthogonal states obtained by measuring $m=3$ particles. The table lists the parameters of the generated state ($r$ and $z$), as well as the experimental parameters of the generation scheme (the squeezing of the input oscillators $\text{S}_1$, $\text{S}_2$ in dB, and the beamsplitter amplitude transmission coefficient $t$).}
\end{table}

\section{Example of generation parameters for optimal codewords} \label{Append_primeri}
\begin{figure}[H]
    \centering
    \includegraphics[width=1.0 \linewidth]{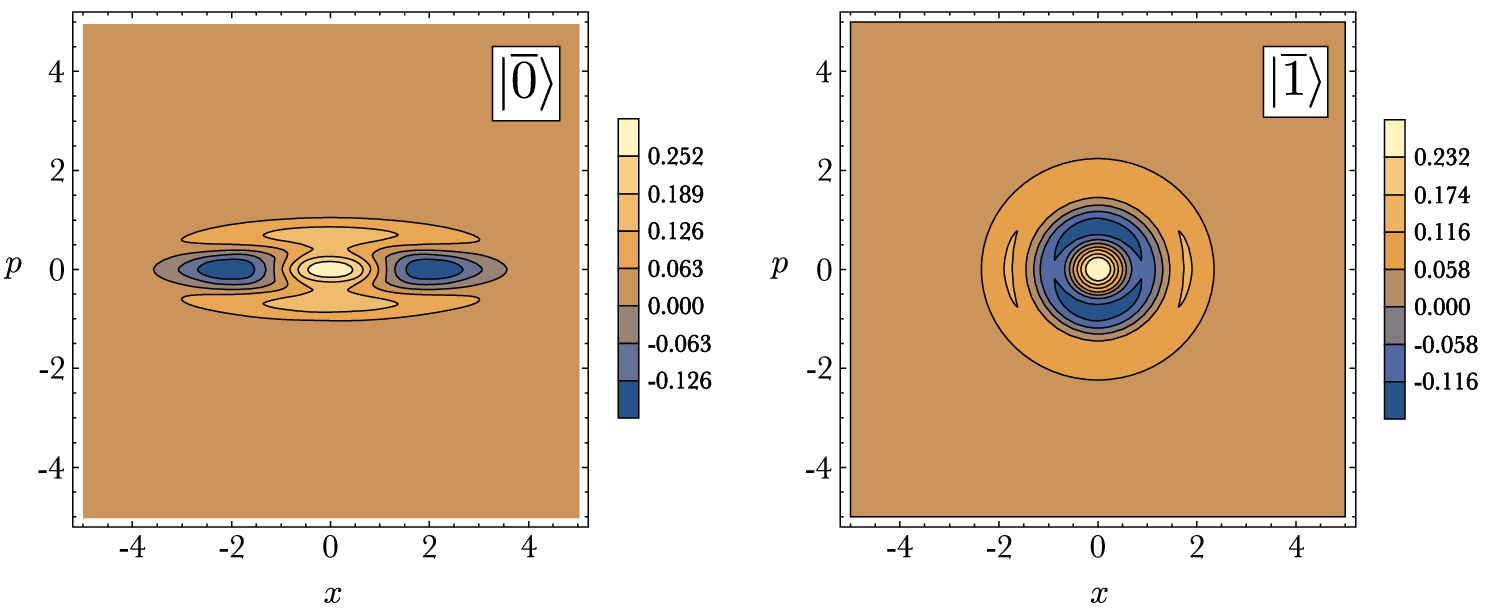}
    \caption{Contour plots of the Wigner functions for two CW with $m=2$, optimal for correcting particle loss errors at a damping parameter of $\gamma=0.01$.}
    \label{fig:wig_count}
\end{figure}

Figure \ref{fig:wig_count} shows the contour plots of the Wigner functions of two CW optimized for correcting particle loss errors with a damping parameter of $\gamma=0.01$. The CW are obtained by measuring two particles, $m=2$.
\noindent From the figure, it is evident that the CW differ significantly from each other. To generate the codeword $|\overline{0}\rangle$ (left frame), the scheme shown in Fig. \ref{fig:scheme} requires the following parameters: $\text{S}_1 \approx 10$ dB, $\text{S}_2 \approx -9.2$ dB, and $t \approx 0.1$, where $\text{S}_i$ denotes the squeezing of the $i$-th oscillator. The codeword $|\overline{1}\rangle$ (right frame) is generated with the scheme parameters: $\text{S}_1 \approx 9.2$ dB, $\text{S}_2 \approx -10.6$ dB, and $t \approx 0.5$.

\section{Codewords rotated by $\pi/2$ on the phase plane} \label{append_pi2}
To understand how to relate the parameters of the CW (\ref{1_CW}) and (\ref{2_CW}) to rotate them by $\pi/2$ on the phase plane, we need to analyze their Wigner functions. The Wigner function of a state generated by the scheme shown in Fig. \ref{fig:scheme} takes the following form:
\begin{align} \label{Wig_funct}
   W_n\left(r,z,x,p\right)= \frac{e^{-e^{-2 r} p^2 - e^{2 r} x^2} (-1)^n}{\pi \, {}_2F_1\left(\frac{1}{2} - \frac{n}{2}, -\frac{n}{2}, 1, z^2\right)}
\sum_{k=0}^n \binom{n}{k} \frac{1}{k!} \left(-\frac{z}{2}\right)^k
H_k\left(\frac{e^r x + i e^{-r} p}{\sqrt{z}}\right)
H_k\left(\frac{e^r x - i e^{-r} p}{\sqrt{z}}\right)
\end{align}
When rotating by an angle $\pi/2$, the Wigner function changes as follows:
\begin{align}
     W_n\left(r,z,x,p\right) \rightarrow  W_n\left(r,z,-p,x\right).
\end{align}
It follows that the CW (\ref{1_CW}) and (\ref{2_CW}) will be rotated by an angle $\pi/2$ provided that the following equality is satisfied:
\begin{align} \label{append_eq}
    W_n\left(r_1,z_1,x,p\right)=W_n\left(r_2,z_2,-p,x\right).
\end{align}
From Eq. (\ref{Wig_funct}) it is clear that condition (\ref{append_eq}) holds for any $m$ when $z_2=-z_1$ and $r_2=-r_1$.

\section{Experimental parameters for generating rotated CW with $m=6$ best suited for correcting dephasing errors} \label{Append_m_6}
\begin{figure}[H]
    \centering
    \includegraphics[width=0.8\linewidth]{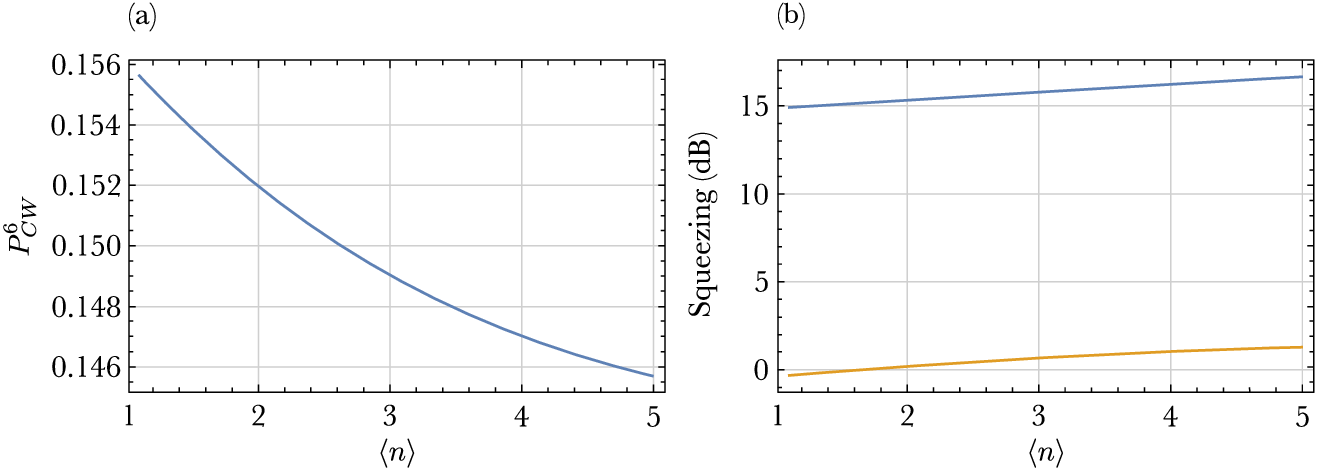}
    \caption{Dependences of the joint probability of generating CW (frame (a)) and the squeezing of the required oscillators (frame (b)) on the mean particle number in the CW. Both plots correspond to CW obtained by measuring $m=6$ particles. In frame (b), different colors indicate the squeezing of the two oscillators required for CW generation.}
    \label{fig:res_n6_deph}
\end{figure}

Let us now estimate the experimental parameters for generating CW obtained by measuring $m=6$ particles, which are best suited for correcting dephasing errors. Fig. \ref{fig:res_n6_deph} (a) shows the joint probability of generating CW as a function of the mean particle number in the CW. Fig. \ref{fig:res_n6_deph} (b) presents the dependence of the required oscillator squeezing for generating optimal CW with maximum probability on the mean particle number.
\noindent From the figure, it can be seen that the probability of generating optimal CW with $m=6$ is only a few tenths of a percent, while the required squeezing of one of the oscillators always exceeds 15 dB.
\end{document}